# Non-Volatile Analog Control and Reconfiguration of a Vortex Nano-Oscillator Frequency

M. Stebliy, A. S. Jenkins, L. Benetti, E. Paz, R. Ferreira

International Iberian Nanotechnology Laboratory, INL, Av. Mestre Jose Veiga s/n, 4715-330, Braga, Portugal

Email Address: maksim.steblii@inl.int

## Abstract

Magnetic tunnel junctions are nanoscale devices which have recently attracted interested in the context of frequency multiplexed spintronic neural networks, due to their interesting dynamical properties, which are defined during the fabrication process, and depend on the material parameters and geometry. This paper proposes an approach to extending the functionality of a standard magnetic tunnel junction (MTJ) by introducing an additional ferromagnet/antiferromagnet (FM/AFM) storage layer (SL) vertically integrated with the standard vortex MTJ stack into the nanopillar. The magnetostatic field created by this storage layer acts on the free layer and can be used to change its static and dynamic properties. To tune the magnitude and direction of this magnetostatic field, magnetic reconfiguration is carried out through a thermally assisted switching mechanism using a voltage pulse that heats the AFM layer in the SL above the Néel temperature in the presence of an external field. It is experimentally shown that using an MTJ based on a 600 nm diameter nanopillar with a vortex in the free layer, reconfiguration of the SL allows to continuously change the core precession frequency in the 15 MHz range, or around 10% of the device frequency. The reconfigurable analogue storage layer locally affects both the static and dynamic properties of the MTJ free layer, demonstrating vertical 3D integration of additional functionalities into a single MTJ nanopillar.

## 1. Introduction

Neuromorphic spintronics is an emerging field, which is garnering significant interest in the context of designing brain-inspired networks composing of individual spintronic elements, such as Magnetic Tunnel Junctions (MTJs). The MTJ makes it possible to convert changes in the orientation of the magnetization into a change in the resistance of the structure [1]. A wide range of potential neuromorphic spintronic paradigms have been investigated [2], including a novel frequency multiplexed approach where MTJs act as both radio-frequency neurons and synapses [3], where the inherent frequency selectivity of the devices allows for information to be encoded via the device operational frequency. For example, if an MTJ neuron was operating at $f_{neuron}$, this signal could be applied simultaneously via a common bus with many other signals at different frequencies, and only the devices in the synaptic layer with a similar device frequency would be activated. This inherent frequency filtering effect is due to the magnetic properties of the free layer of the MTJ and represents one of the key advantages of MTJs in the context of neuromorphic computation.

Traditionally, the magnetic properties of an MTJ are fixed after device fabrication and depend on the device geometry, the MTJ stack and material parameters. This includes the frequency of MTJs used either as artificial synapses or artificial neurons. Consequently, one of the drawbacks of radio-frequency neuromorphic networks is the rigidity of the network once it has been manufactured, which is detrimental for on-chip learning. In previous works, for example [4], synaptic weighting adjustment required for learning has been achieved using an integrated field line which can produce a local in-plane magnetic field to tune the MTJ frequency. Whilst this enables learning and a modification of the synaptic weights, having to continuously apply large DC currents to set each MTJ in a synapse is energetically expensive and non-volatile reprogrammable weighting is essential for the low power implementation of such large scale networks.

A solution that has been proposed to achieve non-volatile reprogrammability in such networks concerns the use of multiple non-volatile MTJs memory elements in resistive networks [5] [6], however this raises concerns over the scalability of such solution in terms of overall footprint.

A preferred solution would be, therefore, to integrate the storage capacity into the nanopillar itself, such as the use of intermediate magnetic states [7], the proposed reconfigurable storage layer [8] or changes in magnetic parameters [9], which has been previously explored via micromagnetic simulations. But in the specific case of spintronic neuromorphic networks, such non-volatile mechanisms still have to be integrated with MTJ elements acting as RF synapses and RF neurons which have quite different functional requirements and are usually implemented with very specific and optimized stacks. It has recently been shown that exploiting internal degrees of freedom of the vortex free layer [19] can partially provide a mechanism to achieve this goal. While manipulating the vortex chirality can be used to manipulate the vortex frequency, this mechanism is binary (only two possible states can be encoded) and the change in frequency is very limited.

The ideal solution to this problem, which is demonstrated in this paper, is the vertical integration of an analog non-volatile memory layer together with an MTJ stack optimized to act as an artificial RF neuron or artificial RF synapse. It is shown that the dynamic properties of an MTJ with a magnetic vortex in the free layer can be modified by the stray magnetic field created by an additional storage layer (SL) composed by a ferromagnet/antiferromagnet located at a distance of 10 nm above the free layer, as show in Fig. 1. The magnitude and direction of the stray field acting on the vortex free layer can in turn be modified by changing the magnetic configuration of the storage layer. The reconfiguration of the magnetic storage layer is made by applying a large voltage pulse to the MTJ nanopillar in the presence of an external magnetic field. When the voltage pulse amplitude and duration is large enough for the resulting Joule heating to bring the temperature of the antiferromagnet in the storage layer above the Néel temperature, the external magnetic field sets a new minimum energy configuration of the ferromagnet in the SL which depends on the magnitude and direction of the external field. This configuration is then frozen once the voltage pulse subsides and the storage layer cools down below the Neél temperature of the AFM2 antiferromagnet in the presence of the external magnetic field. Once the external magnetic field is removed, the new configuration remains in place indefinitely due to the exchange coupling between ferromagnet and AFM2. As discussed in [3], the weighting of an RF synapse is proportional to the difference between the incoming signal and the signal of the RF synapse ($f_{source}$ - $f_{MTJ}$), so by reprogramming the dynamic properties of the MTJ device, the weight can be reprogrammed in a non-volatile manner. Additionally, it is show that the final frequency of operation of the device can be manipulated in an analog way, i.e., the frequency can be programmed continuously in a given range taking profit of non-trivial magnetic configurations of the storage layer.

Whilst an external magnetic field is still required to reconfigure the storage layer, it is crucially only required during the learning phase when the synaptic weights and frequencies of artificial synapses and neurons are being adjusted to tackle a specific computation, and this can be done with very fast voltage pulses that massively reduce the energy consumption during the learning phase compared to a volatile weighting equivalent. Furthermore, upon the initial programming, this weighting mechanism does not require any additional energy to be kept, meaning that is has zero energy consumption during the inference phase of a neuromorphic network.

This new type of device, acts as a non-volatile frequency programmable oscillator (NV-FPO) and has extremely promising functional properties for the implementation of artificial synapses and artificial neurons in the context of radio-frequency neuromorphic networks or for the implementation of programmable DC->RF and RF->DC transducers in the context of communication systems.

Furthermore, this new mechanism enables a fine-control of critical device properties upon fabrication. Devices implemented with magnetic tunnel junctions depend critically of specific geometrical parameters such as the free layer thickness and nano-pillar diameter. Minute changes in these dimensions can give rise to strong device-to-device variation of key properties and performance, which is particularly detrimental in systems that make use of multiple MTJs. The non-volatile reprograming mechanism reported here can be used to fine-tune key device properties after nanofabrication, making it easy to set up networks of devices in systems that depend critically on the specific properties of each individual device, such as neuromorphic computation networks, memory-in-logic arrays or Wheatstone bridges.

## 2. Investigated Structure

The proposed approach was experimentally studied in an MTJ nanopillar with a diameter of 600 nm and the following composition of the magnetic stack IrMn(15)/CoFe(2)/Ru(0.8)/CoFeB(2.6)/MgO/CoFeB(2)/Ta(0.2)/NiFe(7)/Cu(10)/NiFe(6)/IrMn(6), where thickness are indicated in nm. A schematic representation of the structure is shown in the Fig.1a. The top NiFe/IrMn bilayer performs the function of a storage layer, and the Cu is used as a spacer layer to separate and magnetically decouple it from the free layer. The MTJ has a TMR ratio of 100% and resistance×area product 10 $\Omega\mu m^2$ [10]. The details of preparation can be found in the Materials and Methods section, and magnetic characterisation in the S.1 (at the end of the article).

The orientation of the magnetization in the reference and storage layers is fixed due to the effect of the exchange bias at the interface with the antiferromagnet AFM1 and AFM2, respectively. Reconfiguration requires heating the SL above the Néel temperature of AFM2, which differs from that of the antiferromagnet pinning the synthetic antiferromagnet (SAF) reference layer (AFM1) due to the different thicknesses of the IrMn layers [11] [12]. Using a continuous film, the Néel temperature for the two AF layers was determined to be $T_N^{AFM1}$ = 230 °C and $T_N^{AFM2}$ = 180 °C. The thicknesses for the two antiferromagnets where chosen to make it possible to heat the storage layer (AFM2) above the Néel temperature while keeping the reference SAF pinned. These measurements are presented in more detail in the supplementary materials, S.2. The resulting Néel temperature estimate is an approximation, as the temperature may decrease as the film is patterned [13].

The reconfiguration of the storage layer requires the application of a voltage pulse, $U_{REC}$, between the bottom and top electrodes in the presence of an external field $B_{REC}$. For a sufficiently strong and long voltage pulse, the AFM2 layer can be heated above the Néel temperature, which unlocks the magnetization of the SL. Then the spin configuration of the storage layer relaxes in accordance with the orientation and strength of the $B_{REC}$. At the end of the pulse, the structure cools down, and the storage layer magnetic configuration freezes. The presence of a pinned magnetic vortex in the storage layer suggests indirectly that the current magnetic state of the storage layer is imprinted on the antiferromagnetic layer, i.e. that the antiferromagnet has a non-homogeneous domain structure. Magnetic reconfiguration leads to a change in the magnetostatic stray field $B_{MS}$ created by the SL and acting on the free layer.

The external field applied by the electromagnet for sweeping when constructing hysteresis loops or creating a constant bias is designated as $B_{EX}$. There are also two internal magnetic fields acting on the free layer: the magnetostatic one created by the storage layer $B_{MS}$, and the field of interaction with the pinning layer $B_{PIN}$. All these fields are oriented along x-axis, unless otherwise noted. The magnetic configuration of the MTJ free layer depends on the external magnetic field $B_{EX}$. Under small external magnetic fields the magnetic ground state will be a magnetic vortex as shown by the characteristic magnetic hysteresis loop with a clear intermediate resistance state, Fig.1c. The magnetic vortex in the free layer can be characterized by its natural frequency, which is associated with the orbiting gyration of the vortex core [14] and is determined by the composition and geometry of the disk [15]. However, in the presence of a $B_{EX}$ field, the vortex core will become displaced from the centre and there will be a variation in the resonant frequency [16]. In the structure under study, instead of an external field, a variable stray field of the storage layer is used to this effect.

3. Results

At the first stage, the parameters necessary for the reconfiguration of the storage layer were determined, and the range of $B_{MS}$ that can be created by the storage layer was evaluated. At the second stage, the relationship between the value of storage layer $B_{MS}$ and the resonant frequency of the free layer vortex was determined. At the third stage, micromagnetic simulations were used to reproduce the experimental results and validate the assumptions made about the processes occurring in the system.

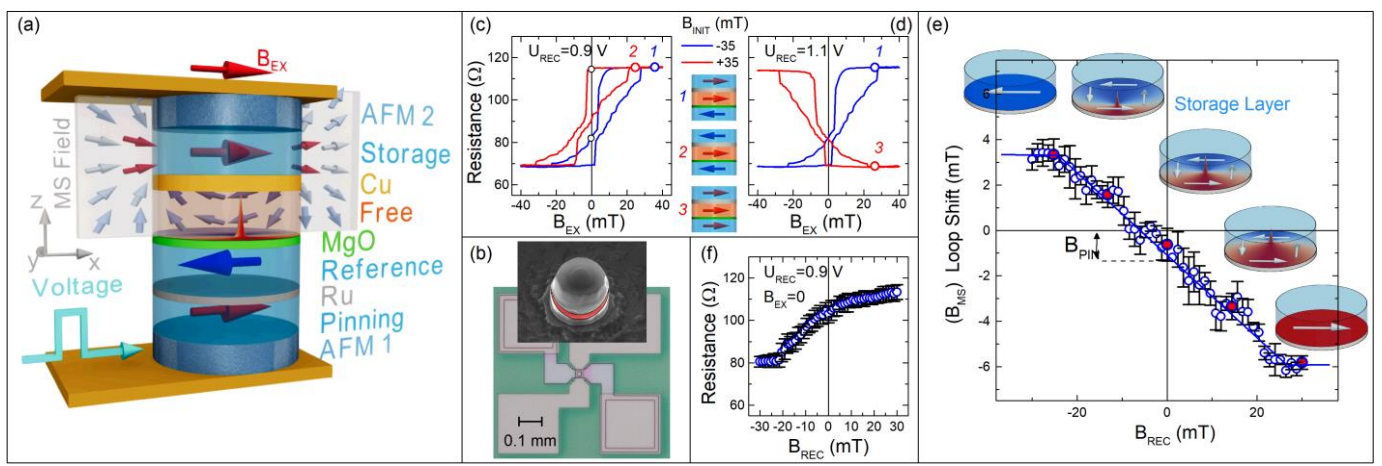

Fig.1. (a) Schematic representation of the structure under study, which indicates the position of the functional layers and external influences. The grey box shows a cross-section of the distribution of the magnetostatic field

*produced by the storage layer and acting on the free layer. (b) Visualization of the structure using an optical microscope and a scanning electron microscopy image of the column at the base of this structure. The red zone indicates the location of the magnetic stack. Comparison of the hysteresis loop obtained after pulse of 0.9 V (c) and 1.1 V (d) with a duration of 1 ms in the presence of a reconfiguartion field of ±35 mT. The sketches show the mutual orientation of the magnetization in the storage, free and pinned layer for the corresponding state on the loops. Dependence of the loop shift (e) and the zero field resistance (f) on the value of the reconfiguration field for a pulse with fixed parameters of 0.9 V and 1ms. The dependences was obtained by averaging ten scans. The insets in figure (e) describe the corresponding magnetic structure of the storage layer.*

### 3.1. Non-volatile reconfiguration of static properties

When a voltage pulse of fixed duration is applied to the devices under a strong $B_{REC}$ field, several outcomes are possible. For voltage pulses with an amplitude lower than 0.8 V the Néel temperature of the storage layer is not reached and the magnetic configuration remains unchanged. Above this threshold, two outcomes are still possible depending on the pulse amplitude, as illustrated in fig 1 c) and d) for pulses with a duration of 1ms and a $B_{REC}$ field of 35mT applied along the easy axis of the SAF reference layer.

For a relatively smaller voltage pulse, i.e. $U_{REC}$ = 0.9 V, a horizontal shift in the hysteresis loop is observed, as shown in Fig.1c. This occurs as a result of the voltage induced Joule heating resulting in temperature larger than AFM2 but less than AFM1, and as such only the orientation of the magnetization of the storage layer is altered: from the state (1) to (2) depending on the sign of the field sense during the reconfiguration pulse, i.e. $B_{REC}$ = -35 or +35 mT, respectively. By increasing the voltage further, i.e. $U_{REC}$ = 1.1 V, the Néel temperature of both layers AFM2 and AFM1 can be reached. In this case, when $B_{REC}$ is applied along the direction and sense of the original post-deposition annealing process field, i.e. $B_{REC}$ = -35 mT, there is no substantial change in the hysteresis loop (i.e. it remains in state 1). However, when the sense of the field applied during reconfiguration is reversed, i.e. $B_{REC}$ = +35mT, the hysteresis loop is completely reversed as both AFM1 and AFM2 have been reset, i.e. state (3), Fig.1d. With such a change, the orientation of the magnetization in the free layer must also be reversed to obtain the minimum resistance of the MTJ. In order to consistently study the effect of switching the storage layer, it is necessary to avoid switching the reference layer. In order to use the devices as intended, a fundamental requirement is that there must exist an operational space for $U_{REC}$ for which the Néel temperature of AFM2 can be reached (allowing the reconfiguration of the storage layer) but the Néel temperature of AFM1 is not reached (keeping the reference SAF undisturbed). This requirement demands a careful choice of the materials and thicknesses of the two anti-ferromagnets in the stack. A detailed characterization of this operational window is made in the supplementary material S.3. For the remaining results reported in this paper, a duration of 1 ms and amplitude 0.9 V of the reconfiguration voltage pulse were kept fixed.

The magnetostatic field acting on the free layer was estimated by measuring the shift of the hysteresis loop, which describes the change in the MTJ resistance when the $B_{Ex}$ field is swept in a direction collinear with the easy axis of the SAF reference layer.

For a fixed voltage pulse, the external magnetic field applied during reconfiguration, $B_{REC}$, was swept in the range from -35 to +35 mT. The obtained dependence of the transfer curve shift has a saturation for $B_{REC}$ exceeding ±23 mT, as shown in Fig.1e, indicating that this is the field amplitude required to saturate the storage layer ferromagnet during the reconfiguration step. In this case, the shift varies in the range from +3.2 to -5.8 mT. Since the magnitude of the loop shift corresponds to the magnitude of the magnetostatic field, it is possible to estimate the amplitude of the $B_{MS}$ equal to 4.5 mT. It should be noted that the transfer curve shift is asymmetric with respect to the zero value of the $B_{REC}$. This happens due to interaction between the free layer and the with SAF reference layer, which may include magnetostatic, indirect exchange or Néel coupling and is described hereinafter in terms of the "effective magnetic field" -$B_{PIN}$, equal to -1.2 mT. To confirm this assumption, it was shown that the effective field changes to +1.2 mT when the pinned layer freezes in the opposite direction, S.4. The dependence of the $B_{MS}$ on the reconfiguration field was repeated ten times and the Fig.1e shows the data with averaging and standard deviation. The noisiness of the dependence is due to the instability of the nucleation of the vortex core, S.6.

Between saturation regions, the shift changes gradually as a function of $B_{REC}$, which indicates that in this range the ferromagnetic state of the SL is not fully saturated. In fact, these results indicate that in this region the relatively thick ferromagnetic layer of the SL is in a vortex state with different values of the core displacement frozen in the storage layer. A detailed discussion of the magnetostatic field generated by the vortex structure is presented in the section 3.3. As a result, the magnetic reconfiguration of the storage layer can continuously change the $B_{MS}$ acting on the free layer in the range of ±5 mT. In accordance with Fig.1c, it can be seen that the shift is reflected in the resistance of the structure in the absence of $B_{EX}$. The dependence of the MTJ resistance on the value of the reconfiguration field is shown on the Fig.1f, which demonstrates that the reconfiguration procedure can be used to program the zero field resistance of the structure.

*3.2. Non-volatile reconfiguration of dynamic properties*

A change in the magnetostatic field changes not only the static parameters of the system, but also the dynamic ones, and more specifically, the resonant frequency of the vortex core precession in the free layer. The spin-torque diode effect was used to experimentally measure this effect [17] [18] [19]. Using a bias-tee, the system is excited by an RF signal and the resulting DC voltage is measured in the absence of any external field. The frequency is swept making it possible to construct a spectrum dependence that gives the value of the resonant frequency by monitoring the DC voltage across the device. Fig.2a shows a set of spectral dependencies for different powers of the input RF signal with a fixed state of the storage layer. This case corresponds to a single-domain state of the storage layer. Reconfiguration in field $B_{REC}$=-6 mT leads to the formation of a magnetic vortex in the storage layer and, consequently, to a significant decrease in the magnetostic field acting on the free layer, Fig.1e. The consequence of this is a decrease in the resonant frequency of the gyrotropic motion of the vortex in the free layer, Fig.2b. Reconfiguration in field $B_{REC}$=+6 mT leads to a further shift of the set of spectra towards lower frequencies, Fig.2c.

Considering a fixed frequency in the presented examples (135 MHz - the area highlighted in green), it is possible to plot the dependences of the system's response on the power of the alternating signal for different states of the storage layer, Fig.2d. The resulting set of dependences shows that the described structure can convert an RF input signal into a DC output signal proportional to power, which is the functionality of an RF synapse [3], where the MTJ multiplies the RF input by a constant, equivalent to a synaptic weight, which depends on the difference between the input signal frequency and the junction's resonant frequency ($f_{source} - f_{MTJ}$). Whilst this has previously been achieved by applying volatile magnetic fields, in Fig2.d the weighting is achieved in a non-volatile manner by reconfiguring the storage layer, demonstrating a non-volatile RF synapse.

In order to illustrate the range of resonant frequencies available for change, the spectra for $B_{REC}$ from -30 to +30 mT were combined into a diagram, Fig.2e. The presented sweep, obtained at a fixed power -5 dBm, contains one hundred steps and each step was repeated ten times to estimate the standard deviation. It can be noted that the resonant frequency sharply decreases in the vicinity of $B_{REC}$=-20 mT, which is associated with the nucleation and freezing of a vortex in the storage layer, which leads to a sharp decrease in the magnetostatic field. With a further $B_{REC}$ decrease, the vortex core is fixed closer and closer to the center of the storage disk, which leads to a smooth decrease in the resonant frequency. The frequency can be changed in the range from 142 to 125 MHz. However, the minimum frequency is characterized by maximum noise, which is presumably due to the poor repeatability of the vortex relaxation process in the vicinity of the storage disk center. The dependences shown in Fig.2e was obtained for the case of the same chirality of vortices in the storage layer and the free layer. Changing mutual chirality can slightly change the dependencies, S.5. As with the static resistance change shown in fig. 1e, there is an associated error during the reconfiguration process, however, this error is less than the total variation in frequency and resistance and can is believed to be mostly associated with the impact of the granularity on the precise vortex core position during reconfiguration. Future devices using amorphous materials should significantly improve this error [20].

In the given example, oscillations can be excited only in the negative region of the reconfiguration fields. This behaviour is associated with the peculiarities of excitation of the spin-torque diode effect. The external RF signal induces vortex precession mainly due to the alternating Oersted field [21]. Such a field does not lead to excitation in the case of a symmetrical arrangement of the vortex. The insert Fig.2f. schematically shows green areas for the vortex core location where oscillations can be excited, and red areas where they cannot. The presence of the $B_{PIN}$ from the pinning layer displaces the vortex core relative to the centre. In cases of negative $B_{REC}$, the core moves within the green zone in region -y. In the case of positive $B_{REC}$, the core moves towards the center, while the field magnitude of $B_{MS}$ is not enough to enter the green zone in region +y. Also, this lack of signal in positive fields may be partly due to the small amplitude of the oscillations, as predicted by micromagnetic simulation, S.5. If the magnetization orientation in the pinning layer changes, the situation will change to the opposite, as detailed in supplementary information S.4.

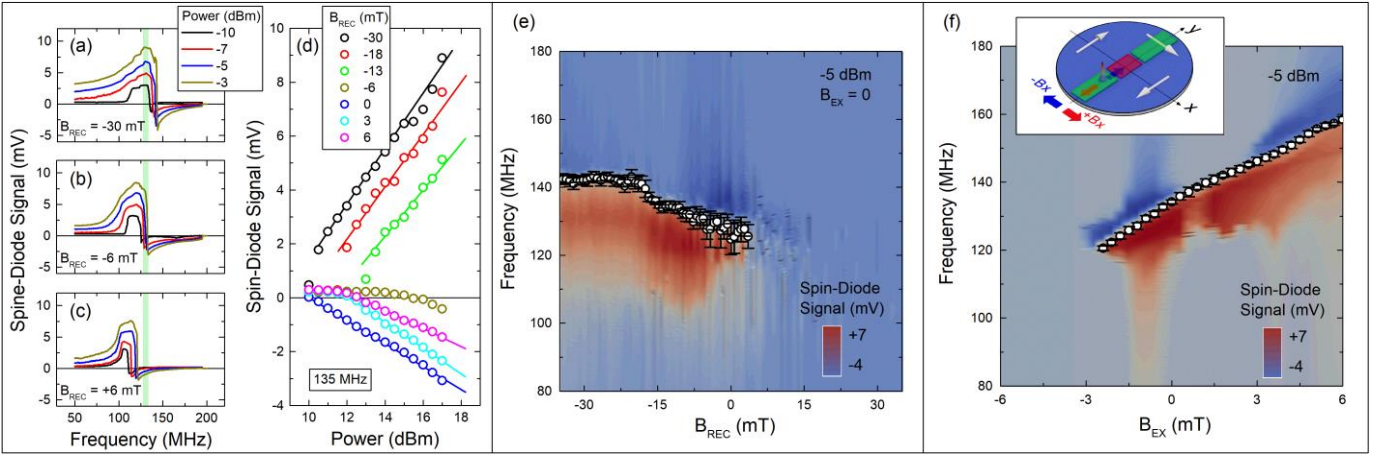

*Fig.2. (a-c) The sets of spectra obtained as a result of the spin-diode effect at different RF signal powers for three different magnetic states of the storage layer. A certain frequency is marked in green for further analysis. (d) Dependence of the system response at a fixed frequency of 135 MHz on the power of the RF signal for different states of the storage layer. (e) The diagram was obtained by combining spectra at -5 dBm for different reconfiguration fields. The dots indicate the resonant frequency. The dependence was obtained by averaging ten scans. (f) Spin-diode frequency versus $B_{EX}$, for a configuration of the storage layer that generates an average zero field in the free layer. Insert demonstrates scheme of the location of the vortex core in the disk in the absence of an external field and a field from the storage layer. The region where vortex oscillation can be excited is marked in green, and in red where it cannot.*

To confirm the assumption that the changes in frequency are being caused by a variation of the internal $B_{MS}$ field, a comparison was made with the action of the external $B_{EX}$ field. For this experiment, the $B_{MS}$ from the storage layer was minimised by promoting the formation of symmetric vortex in the storage layer (i.e. vortex core located at the centre of the disk) that does not create a magnetostatic field (i.e., the shift in the transfer curve loop is equal to $B_{PIN}$ alone. After setting this configuration, the spin-torque diode effect curves were measured for field values in the range $B_{EX}=\pm6mT$, which corresponded to the motion of the vortex core along the y-axis without leaving the disk. The resonant frequency dependence of the spin diode curves obtained are summarized on the diagram of Fig.2f. The increase in the spread of the frequency range from 122 to 158 MHz is associated with an increase in the amplitude of a magnetic field acting on the free layer, compared with the amplitude of $B_{MS} = \pm4.5$ mT achievable by reconfiguring the storage layer between the two saturated states. The presented field sweep was also repeated ten times to evaluate the noise. It can be seen that in the absence of a reconfiguration process, the behavior of the system is much more repeatable.

### 3.3. Micromagnetic simulations

To provide support to the analysis of the experimental data, simulations were carried out in MuMax3 [22]. The goal of the simulations was to replicate the dependence of the vortex free layer resonant frequency on the storage layer magnetic configuration set by $B_{REC}$. To this end, the simulation followed the following steps: for each value of $B_{REC}$, the ground state of the ferromagnet in the storage layer was found; ii) from this configuration, the stray field $B_{MS}$ acting on the free layer was computed and iii) the resonant frequency of the free layer under the action of $B_{MS}$ was extracted. This strategy simplifies the problem at hand by neglecting the effect of the free layer stray field on the storage layer (whose magnetisation is considered fixed). The effect of the reference SAF on the free layer is accounted for by introducing in the free layer and additional constant magnetic $B_{PIN}=+1.5mT$ directed along axis x. The simulation code is given in S.10. The Fig.3a. shows an example of a displaced vortex frozen in the storage layer for a particular value of $B_{REC}$. In this case, the vortex in the free layer is shifted from the center as a result of the action of $B_{MS}$. The $B_{MS}$ distribution acting on the free layer is shown in the grey cylinder. One hundred files with the magnetostatic field distribution were prepared in the same way for the cases of different values of $B_{REC}$ acting on a storage layer starting from a vortex in the centre of the layer.

In the considered example of Fig.3a, the vortex core in the storage layer is displaced in +y direction, and, as a result most of the disk is magnetized along +x, and only a small part along -x. Nevertheless, the magnetostatic field at all points has a component directed along -x. To change the direction of the magnetostatic field to positive, the core of the vortex must be displaced into the -y region. A non-uniform field can be seen under the core and a non-zero z-component at the edge, show also in Fig.3b. The field will also be non-uniform in the case of a single-domain state of the storage layer, discussed further in the supplementary materials S.7. Despite the inhomogeneity, the magnetostatic field can be characterized by a certain average value $B_{MS}^x$ which is computes as a function of $B_{REC}$ in Fig.3c. When fields are greater than ±40 mT, the storage layer goes into a single-domain state and the value of $B_{MS}$ reaches saturation. If the core is in the center, then the average value of $B_{MS}$ is close to zero. This dependence matches qualitatively the experimental data shown in Fig.1f.

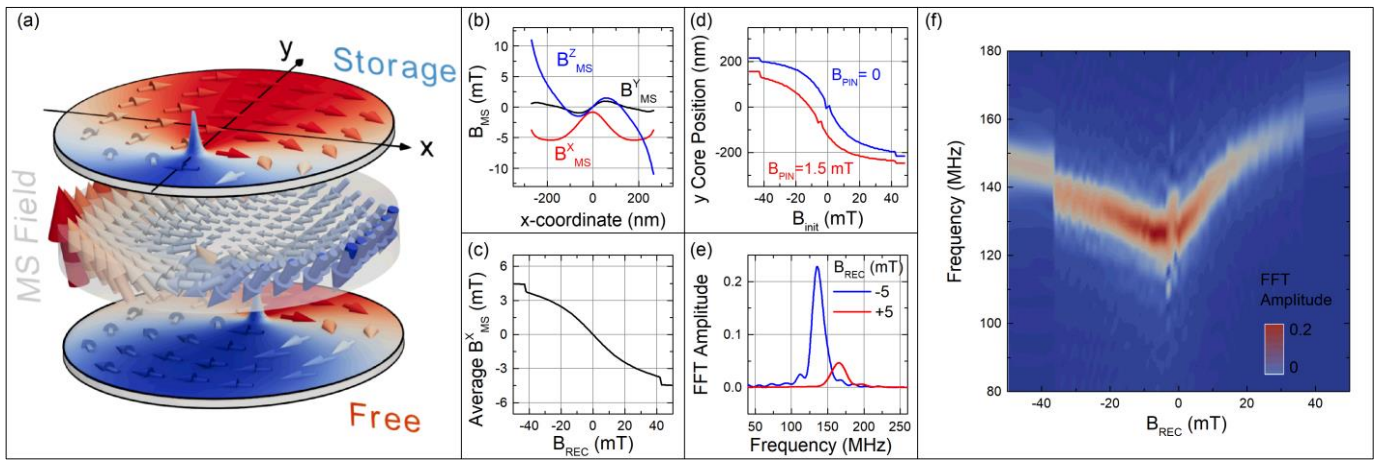

*Fig.3. Results of micromagnetic simulation. (a) An example of the magnetic structure of a free layer for the case of freezing of an asymmetrical vortex in the storage layer without external field. The grey cylinder contains the distribution of the magnetostatic field created by the storage layer in the free layer. The (b) plot shows the dependences of the components of the magnetostatic field in the free layer along the central x-axis for this case. Dependence of the average x-component of the magnetostatic field induced by the storage layer (c) and vortex core position (d) in the free layer on the value of the reconfiguration field. (f) Diagram resulting from the summation of the spectrum obtained with different values of the reconfiguration field. The examples of the spectrum (e).*

The Fig.3d. shows the dependence of the coordinate of the vortex core in the free layer on the $B_{REC}$ field in the presence and absence of $B_{PIN}$. The action of this field leads to a shift in the positional dependence and causes the branches to be asymmetric around zero field. The presence of inhomogeneity in the vicinity of zero is associated with the interaction of the vortex cores, which are vertically aligned. For this study, the chirality and polarity of both vortices was set to be the same.

For each reconfiguration field value that determined the distribution of the magnetostatic field from the storage disk, a resonance curve of the free layer disk was obtained by analysing the relaxation processes after excitation of the system by the pulse, as detailed in supplementary information S.8. Two examples of spectrum obtained for two particular values of $B_{REC}$ resulting in different states of the storage layer are presented in Fig.3e. In Fig. 3f the frequency output of the device for the full range of simulated $B_{REC}$ values is summarized. This figure can be directly compared to the experimental data in Fig2b. The results obtained qualitatively agree with the experimental results. Starting from the left, there is a plateau associated with the presence of the storage layer in a state of saturation. The sharp decrease in frequency at $B_{REC}$=22 mT corresponds to the nucleation of a vortex in the storage layer, then a smooth decrease in the frequency in the range of 10 MHz, similarly to what is observed in the experimental data. Simulation makes it possible to evaluate the oscillations of the vortex, both for positive and negative reconfiguration fields, in contrast to the experiment where the spin-torque diode effect method imposed restrictions. Also, using micromagnetic simulation, the dependence of the resonant frequency on a constant external field was obtained, as was done in the case of Fig.2f. The results obtained are in qualitative agreement with the experiment, S.9.

## 4. Discussion

The qualitative agreement of the result obtained due to experiment and simulation confirms the assumptions made about the processes taking place at the micromagnetic level. The voltage pulse combined with the external field made it possible to reconfigure the storage layer and freeze the single-domain or vortex state with different positions of the core. That approach allows to gradually tune the magnetostatic stray field created by the storage layer and acting upon the free layer in the range from ±5 mT. As result, the frequency of the gyrotropic motion of the vortex in the free layer can be continuously tuned in a 15 MHz range in a non-volatile way. Furthermore, resistance at a fixed value of $B_{EX}$ and rectified voltage at a fixed frequency excitation can also be continuously tuned. Thus, these system parameters can be changed using the reconfiguration procedure, after which their values will be stored non-volatilely as shown in Fig.1f and Fig.2c. These functional characteristics make it possible to implement a programmable resistor, a programmable frequency generator or a programmable frequency rectifier. But these devices have a particularly large potential being used as artificial synapse/neurons with a programmable weight for neuromorphic computation applications.

Reconfiguring the resistance of an MTJ junction can also be used for neuromorphic computing problems. A number of works have experimentally shown that with the help of gradual magnetization reversal of the junction, it is possible to realize artificial synapses that provide a state of multi-weight conductivity. Work [23] uses control of the position of the domain wall in the free layer of a stripline shape MTJ by applying current pulses. In works [24] and [25], an elongated MTJ junction of variable width or with notches is used for step-by-step magnetization reversal under the action of voltage pulses according to spin-orbit torque mechanism. Another example is the gradual magnetization reversal of grains in the MTJ with perpendicular magnetic anisotropy under the action of a series of voltage pulses [26]. In the cases considered, intermediate resistance values are determined by the history of the magnetization reversal process, and it will be lost under the influence of a magnetic field. In the case of using the reconfiguration of the storage layer, the value of the local magnetostatic field will be preserved regardless of the state of the free layer. Also, the approach proposed in the work makes it possible to control the junction resistance at much smaller structure sizes, which is important when scaling down.

The possibility of freezing a magnetic vortex in the storage layer indirectly indicates that a vortex spin structure is also formed in the adjacent antiferromagnet layer. Previously, using direct visualization, it was shown that a vortex [27] [28], antivortex [29], or skyrmion [30] [31] can be imprinted into antiferromagnet. This was observed in both continuous and patterned bilayer FM/AFM films, and both as a result of spontaneous formation and as a result of thermal exposure. In this study, the controlled transition between single-domain and vortex magnetic configurations in the storage layer suggests that the same transition occurs in the antiferromagnet layer states. The ability to reprogram an antiferromagnetic magnetic ground state fully electrically in an integrated MTJ device is a significant achievement for the development of antiferromagnetic spintronics [32].

Although promising, the proposed structure still has room for improvement. Firstly, it should be possible to reduce the duration of the voltage pulses required for heating to the nanosecond range. Presumably, for this it is necessary to increase the pulse amplitude, which is currently limited by the small difference 60 °C in the Néel temperature between AFM1 and AFM2. The difference can be increased by replacing IrMn with FeMn or PtMn in the AFM1 layer [33]. Secondly, to avoid the use of an external reconfiguration field, field-lines can be vertically integrated into the system and placed above the MTJ pillar [4]. Thirdly, to reduce noise in the obtained dependencies, it is necessary to make the process of vortex generation and movement more stable. For this purpose, the polycrystalline NiFe can be replaced for an amorphous ferromagnet like CoFeSiB in the free layer to reduce the interaction of the vortex core with grain boundaries [20] [34] [35].

## 5. Conclusions

It has been experimentally shown that the reconfiguration of an additional AFM/FM storage layer vertically integrated in a standard vortex MTJ makes it possible to continuously change the resonant frequency of the vortex precession in the free layer in a non-volatile manner. A frequency reprogramming range of 15 MHz was demonstrated in a nano-pillar with a diameter of 600 nm. The change is a consequence of the residual displacement of the vortex core in the free layer due to a change in the magnetostatic field created by the storage layer. Also, this displacement changes the resistance of the structure and the magnitude of the response to external alternating excitation. For reconfiguration, voltage pulses with a duration of 1 ms and an amplitude of 0.9 V are used in the presence of an external magnetic field in range ±35 mT. During the pulse the antiferromagnetic layer of the storage layer is heated above the Néel temperature and unlocks the ferromagnetic layer, which reconfigures and freezes after the end of the pulse under the action of the external magnetic field.

6. Material and Methods

The structures under study were obtained based on a film of the following composition from bottom to top:

[Ta(5)/CuN(25)]$_6$/Ta(5)/Ru(5)/Ir$_{20}$Mn$_{80}$(15)/Co$_{70}$Fe$_{30}$(2)/Ru(0.825)/Co$_{40}$Fe$_{40}$B$_{20}$(2.6)/MgO/Co$_{40}$Fe$_{40}$B$_{20}$(2)/Ta(0.21)/NiFe(7)/Cu(10)/NiFe(6)/Ir$_{20}$Mn$_{80}$(6)/Ta(10)/Ru(7)/TiWN(15)/AlSiCu(200)/TiWN(15), thicknesses in nm. The film was deposited on the surface of the 200 mm thermally oxidized wafer Si/SiO$_2$(200 nm) using magnetron sputtering Singulus TIMARIS Multi-Target-Module. The layers below the first IrMn layer are further used to form the bottom electrical contact and are optimized to minimize roughness. The layers above the second IrMn layer are needed to protect the magnetic stack from numerous further etching procedures. Then, nano-pillars with a diameter of 600 nm were formed using electron beam lithography and ion beam milling. The nano-pillar was insulated with SiO$_2$, after which the bottom and top electrical contacts were formed, as shown in Fig.1b. The thickness of the MgO layer was selected to obtain RxA=10 Ωμm$^2$, which maximizes the ouput power of spin-transfer torque nano-oscillators in accordance with previously conducted studies [10]. After fabrication, the structures were annealed at 330 C for 2 hours and the presence of B$_X$ = 1 T, to fix the magnetization orientation in the reference and storage layers due to exchange bias effect at the interface with IrMn.

The magnetic properties of the stack were studied using a vibrating magnetometer (VSM) equipped with a thermal setup using a separate coupon sample with an area of 6x10 mm$^2$. The study of magnetotransport properties was carried out using a probe station equipped with an electromagnet creating an in-plane field with an amplitude of up to 0.1 T and two-point high-frequency probe. To record the hysteresis loop, a Keysight B2901B SMUsource-meter unit (SMU) was used, which measured the resistance of the structure during the magnetic field sweep. To record spin-torque diode effect curves, high-frequency excitation was carried out using the Keysight N5173B Microwave Analog Signal Generator connected through a bias tee. The magnetic reconfiguration procedure was carried out using pulses from the same SMU in the presence of a constant magnetic field created by the electromagnet.

Acknowledgements
This work has received funding from the European Union's Horizon 2020 research and innovation programme under grant agreement No 101017098 (project RadioSpin), No 899559 (project SpinAge) and No 101070287 (project Swan-on-chip).

# Bibliography

[1] Jian-Gang (Jimmy) Zhu, Chando Park, "Magnetic tunnel junctions," *Materials Today,* vol. 9, no. 11, pp. 36-45, 2006.

[2] Grollier, J., Querlioz, D., Camsari, K.Y., "Neuromorphic spintronics," *Nature Electronics,* vol. 3, p. 360, 2020.

[3] Ross, A., Leroux, N., De Riz, A, "Multilayer spintronic neural networks with radiofrequency connections," *Nature Nanotechnology,* vol. 18, p. 1273, 2023.

[4] N. Leroux, A. Mizrahi, R. Ferreira, J. Grollier, "Hardware realization of the multiply and accumulate operation on radio-frequency signals with magnetic tunnel junctions," *Neuromorphic Computing and Engineering,* vol. 1, p. 011001, 2021.

[5] Piotr Rzeszut, Jakub Chęciński, Ireneusz Brzozowski, Sławomir Ziętek, Witold Skowroński, Tomasz Stobiecki, "Multi-state MRAM cells for hardware neuromorphic computing," *Scientific Reports,* vol. 12, p. 7178 , 2022.

[6] T. Böhnert, Y. Rezaeiyan, M. Claro, R. Ferreira, "Weighted Spin Torque Nano-Oscillator System for Neuromorphic Computing," *Communications Engineering,* vol. 2, p. 65, 2023.

[7] Jialin Cai, Bin Fang, Chao Wang, Zhongming Zeng, "Multilevel storage device based on domain-wall motion in a magnetic tunnel junction," *Applied Physics Letters,* vol. 111, p. 182410, 2017.

[8] C.I.L. de Araujo, S.G. Alves, L.D. Buda-Prejbeanu, B. Dieny, "Multilevel Thermally Assisted Magnetoresistive Random-Access Memory Based on Exchange-Biased Vortex Configurations," *Physical Review Applied,* vol. 6, p. 024015, 2016.

[9] Ch. Yun, Y. Wu, Z. Liang, Z. Luo, "Magnetic anisotropy-controlled vortex nano-oscillator for neuromorphic computing," *Condensed Matter Physics,* vol. 10, p. 1019881, 2022.

[10] Costa, J.D., Serrano-Guisan, S., Lacoste, B., "High power and low critical current density spin transfer torque nano-oscillators using MgO barriers with intermediate thickness.," *Scientific Reports,* vol. 7, p. 7237 , 2017.

[11] Christian Rinaldi, Lorenzo Baldrati, Matteo Di Loreto, Marco Asa, Riccardo Bertacco, Matteo Cantoni, "Blocking Temperature Engineering in Exchange-Biased CoFeB/IrMn Bilayer," *IEEE Transactions on Magnetics,* vol. 54, no. 4, pp. 1-7, 2018.

[12] A. J. Devasahayam, M. H. Kryder, "The dependence of the antiferromagnet/ferromagnet blocking temperatureon antiferromagnet thickness and deposition conditions," *Journal Of Applied Physics,* vol. 85, p. 5519, 1999.

[13] L. Lombard, E. Gapihan, R. C. Sousa, Y. Dahmane, Y. Conraux, C. Portemont, C. Ducruet, C. Papusoi, I. L. Prejbeanu, J. P. Nozières, B. Dieny, A. Schuhl, "IrMn and FeMn blocking temperature dependence on heating pulse width," *Journal Of Applied Physics,* vol. 107, p. 09D728, 2010.

[14] V. Novosad, F. Y. Fradin, P. E. Roy, K. S. Buchanan, K. Yu. Guslienko, and S. D. Bader, "Magnetic vortex resonance in patterned ferromagnetic dots," *Physical Review B,* vol. 72, p. 024455, 2005.

[15] Junjia Ding, Gleb N. Kakazei, Xinming Liu, Konstantin Y. Guslienko & Adekunle O. Adeyeye, "Higher order vortex gyrotropic modes in circular ferromagnetic nanodots," *Scientific Reports,* vol. 4, p. 4796, 2014.

[16] V. S. Pribiag, I. N. Krivorotov, G. D. Fuchs, P. M. Braganca, O. Ozatay, J. C. Sankey, D. C. Ralph & R. A. Buhrman, "Magnetic vortex oscillator driven by d.c. spin-polarized current," *Nature Physics,* vol. 3, p. 498, 2007.

[17] A. A. Tulapurkar, Y. Suzuki, A. Fukushima, H. Kubota, H. Maehara, K. Tsunekawa, D. D. Djayaprawira, N. Watanabe, S. Yuasa, "Spin-torque diode effect in magnetic tunnel junctions," *Nature,* vol. 438, p. 339, 2005.

[18] Petr N. Skirdkov, Konstatin A. Zvezdin, "Spin-Torque Diodes: From Fundamental Research to Applications," *Annalen Der Physik,* vol. 532, p. 1900460 , 2020.

[19] L. Martins, A. Jenkins, L. Alvarez, P. Freitas, R. Ferreira, "Non-volatile artificial synapse based on a vortex nano-oscillator," *Scientific Reports,* vol. 11, no. 16094, 2021.

[20] Jenkins, A.S., Martins, L., Benetti, L.C., "The impact of local pinning sites in magnetic tunnel junctions with non-homogeneous free layers," *Communications Materials,* vol. 5, no. 7, 2024.

[21] Flavio Abreu Araujo, Chloé Chopin, Simon de Wergifosse, "Ampere–Oersted field splitting of the nonlinear spin-torque vortex oscillator dynamics," *Scientific Reports,* vol. 12, p. 10605, 2022.

[22] Arne Vansteenkiste, Jonathan Leliaert, Mykola Dvornik, Mathias Helsen, Felipe Garcia-Sanchez, Bartel Van Waeyenberge, "The design and verification of MuMax3," *AIP Advances,* vol. 4, p. 107133, 2014.

[23] S. Siddiqui, S. Dutta, A. Tang, L. Liu, C. Ross, M. Baldo, "Magnetic Domain Wall Based Synaptic and Activation Function Generator for Neuromorphic Accelerators," *Nano Letters,* vol. 20, no. 2, p. 1033, 2020.

[24] T. Leonard, S. Liu, M. Alamdar, H. Jin, C. Bennett, J. Incorvia, "Shape-Dependent Multi-Weight Magnetic Artificial Synapses for Neuromorphic Computing," *Advanced Electronic Materials,* vol. 8, no. 12, 2022.

[25] S. Liu, T. Xiao, C. Cui, M. Marinella, "A domain wall-magnetic tunnel junction artificial synapse with notched geometry for accurate and efficient training of deep neural networks," *Applied Physics Letters,* vol. 118, no. 20, p. 202405, 2021.

[26] D. Kim, W. Yi, J. Choi, K. Ashiba, J. Park, "Double MgO-Based Perpendicular Magnetic Tunnel Junction for Artificial Neuron," *Front Neuroscience,* vol. 14, p. 309, 2020.

[27] G. Salazar-Alvarez, J. J. Kavich, J. Sort, J. Nogués, P. Gambardella, "Direct evidence of imprinted vortex states in the antiferromagnet," *Applied Physics Letters,* vol. 95, p. 012510, 2009.

[28] J. Wu, D. Carlton, J. S. Park, Y. Meng, J. Bokor, Z. Q. Qiu, "Direct observation of imprinted antiferromagnetic vortex states in CoO/Fe/Ag(001) discs," *Nature Physics,* vol. 7, p. 303, 2011.

[29] J. Li, A. Tan, K. W. Moon, A. Doran, C. Hwang, Z. Q. Qiu, "Imprinting antivortex states from ferromagnetic Fe into antiferromagnetic NiO in epitaxial NiO/Fe/Ag(001) microstructures," *Applied Physics Letters,* vol. 104, p. 112407, 2014.

[30] K.G. Rana, R.L. Seeger, S. Ruiz-Gomez, V. Baltz, O. Boulle, "Imprint from ferromagnetic skyrmions in an antiferromagnet via exchange bias," *Applied Physics Letters,* vol. 119, p. 192407, 2021.

[31] Y. Guang, I. Bykova, Y. Liu, X. Han, G. Schütz, "Creating zero-field skyrmions in exchange-biased multilayers through X-ray illumination," *Nature Communicattions,* vol. 11, p. 949, 2020.

[32] T. Jungwirth, X. Marti, P. Wadley, J. Wunderlich, "Antiferromagnetic spintronics," *Nature Nanotechnology,* vol. 11, p. 231, 2016.

[33] V. Baltz, A. Manchon, M. Tsoi, T. Moriyama, T. Ono, and Y. Tserkovnyak, "Antiferromagnetic spintronics," *Reviews of Modern Physics,* vol. 90, p. 015005, 2018.

[34] Shreya, S., Jenkins, A.S., Rezaeiyan, Y., "Granular vortex spin-torque nano oscillator for reservoir computing," *Scientific Report,* vol. 13, p. 16722, 2023.

[35] R. L. Compton, T. Y. Chen, and P. A. Crowell, "Magnetic vortex dynamics in the presence of pinning," *Physical Review B,* vol. 81, p. 144412, 2010.

[36] Viola Krizakova, Manu Perumkunnil, Sébastien Couet, Pietro Gambardella, Kevin Garello, "Spin-orbit torque switching of magnetic tunnel junctions for memory applications," *Journal of Magnetism and Magnetic Materials,* vol. 562, p. 169692, 2022.

[37] A.N. Grigorenko, D.J. Mapps, "Magnetic tunnel junctions with non-collinear anisotropy axes for sensor applications," *Journal of Physics D: Applied Physics,* vol. 36, no. 7, 2003.

[38] Y. C. Lee, C. T. Chao, L. C. Li, Y. W. Suen, Lance Horng, Te-Ho Wu, C. R. Chang, J. C. Wu, "Magnetic tunnel junction based out-of-plane field sensor with perpendicular magnetic anisotropy in reference layer," *Journal of Applied Physics,* vol. 117, no. Magnetic tunnel junction based out-of-plane field sensor with perpendicular magnetic anisotropy in reference layer, p. 17A320, 2015.

[39] A. Dussaux, B. Georges, J. Grollier, V. Cros, A.V. Khvalkovskiy, A. Fukushima, M. Konoto, H. Kubota, K. Yakushiji, S. Yuasa, K.A. Zvezdin, K. Ando, A. Fert, "Large microwave generation from current-driven magnetic vortex oscillators in magnetic tunnel junctions," *Nature Communications,* vol. 1, no. 8, 2010.

[40] M.E. Stebliy, A.G. Kolesnikov, M.A. Bazrov, A.S. Samardak, "Current-Induced Manipulation of the Exchange Bias in a Pt/Co/NiO Structure," *ACS Appl Mater Interfaces,* vol. 13, p. 42258, 2021.

[41] Takeshi Kat, Satoshi Iwata, Daiki Oshima, "Progress on efficient current-induced magnetization switching," *Electrical Engineering In Japan,* vol. 212, no. 1, p. 3, 2020.

[42] M. E. Stebliy, S. Jain, A. G. Kolesnikov, A. V. Ognev, A. S. Samardak, A. V. Davydenko, E. V. Sukovatitcina, L. A. Chebotkevich, J. Ding, J. Pearson, V. Khovaylo & V. Novosad, "Vortex dynamics and frequency splitting in vertically coupled nanomagnets," *Scientific Reports,* vol. 7, no. 1127, 2017.

[43] Jonathan Z. Sun, Christopher Safranski, Philip Trouilloud, Christopher D'Emic, Pouya Hashemi, and Guohan Hu, "Stochastic magnetic tunnel junction with easy-plane dominant anisotropy," *Physical Review B,* vol. 107, p. 184433, 2023.

[44] S. Bandiera, R. C. Sousa, M. Marins de Castro, C. Ducruet, C. Portemont, S. Auffret, L. Vila, I. L. Prejbeanu, B. Rodmacq, B. Dieny, "Spin transfer torque switching assisted by thermally induced anisotropy reorientation in perpendicular magnetic tunnel junctions," *Applied Physics Letters,* vol. 99, p. 202507, 2011.

[45] Gerardin, O., Gall, H., Vukadinovic, N., "Micromagnetic calculation of the high frequency dynamics of nano-size rectangular ferromagnetic stripes," *Journal of Applied Physics,* vol. 89, p. 7012, 2001.

[46] "Magnetic Oscillations Driven by the Spin Hall Effect in 3-Terminal Magnetic Tunnel Junction Devices," *Physical Review Letters,* vol. 109, p. 186602, 2012.

[47] C. T. Chao, C. Y. Kuo, Lance Horng, M. Tsunoda, M. Takahashi, J. C. Wu, "Current-induced switching of exchange bias in nano-scaled magnetic tunnel junctions with a synthetic antiferromagnetic pinned layer," *Journal of Applied Physics,* vol. 111, p. 07B103, 2012.

# Supplementary

## Non-Volatile Analog Control and Reconfiguration of a Vortex Nano-Oscillator Frequency

Maksim Stebliy, Alex Jenkins, Luana Benetti, Elvira Paz, Ricardo Ferreira

International Iberian Nanotechnology Laboratory, INL, Av. Mestre Jose Veiga s/n, 4715-330, Braga, Portugal

Email Address: maksim.steblii@inl.int

Keywords: spintronics, magnetic tunnel junctions, MTJ, magnetic vortex, vortex oscillation, spin-diode effect, thermal assistant switching.

### S.1. Magnetic properties of the film

To form and fix a single-domain state in the storage and reference layers, the multilayer film was annealed for two hours at a temperature of 330 C and a field of 1 T in the plane, Fig.S1a. The Fig.S1b shows an example of a hysteresis loop for a film obtained in a vibration sample magnetometer (VSM) at room temperature. The key stages of magnetization reversal are marked on the loop, illustrated in Fig.S1c. In states (1), the magnetization in all layers is oriented along the external negative field. During the transition from state 1 to state (2), a rotation of the magnetization orientation in the pinned SAF layer occurs under the action of the exchange bias field at the interface with the AFM1 and indirect antiferromagnetic exchange interaction with the SAF reference layer, which has a larger magnetic moment. A further decrease in the field leads to switching of the storage layer (3) under the influence of the exchange bias from the AFM2. An increase in the field in the positive direction leads to switching of the free layer (4), whose loop is shifted due to weak interaction with the pinning layer, that include indirect exchange or Néel coupling and is described in terms of the "effective magnetic field" equal to ±1.5 mT. A further increase in the field will lead to a reversal of the magnetization in the SAF reference layer (5), which is prevented by indirect exchange interaction with the SAF pinned layer.

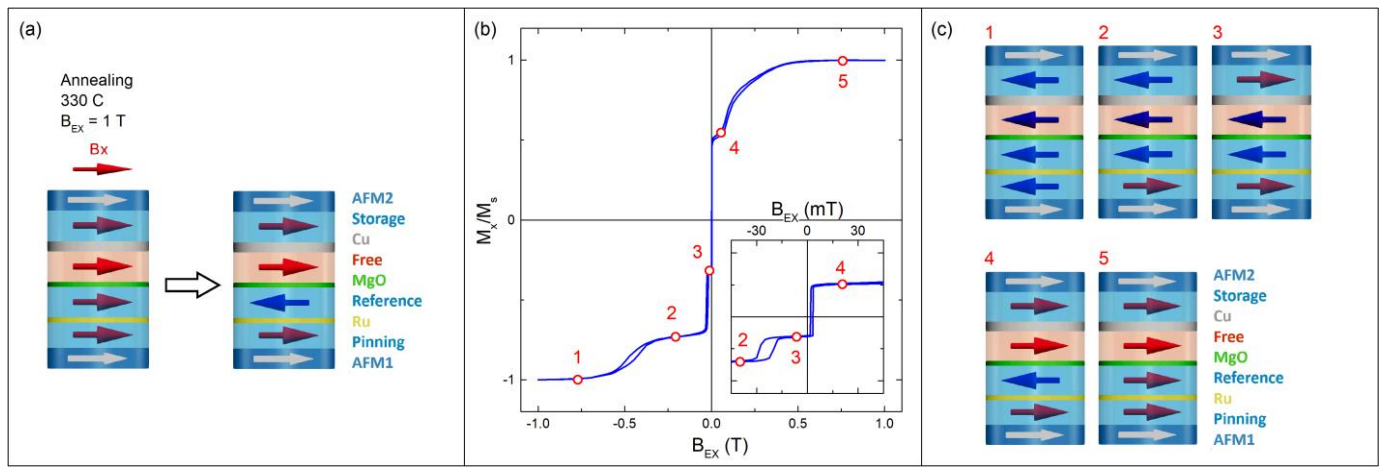

Fig.S1. (a) Scheme of sample annealing and the corresponding residual state after cooling in the field. (b) Magnetic hysteresis loop of the IrMn(15)/CoFe(2)/Ru(0.8)/CoFeB(2.6)/MgO/CoFeB(2)/Ta(0.2)/NiFe(7)/Ta(10)/NiFe(6)/IrMn(6) composition film obtained by the VSM method. The inset shows an enlarged section of the central part of the loop. (c) Scheme of magnetization orientation in layers at key stages of film magnetization reversal.

### S.2. Néel temperature estimation

The structure under study contains two antiferromagnetic layers. Changing the frequency properties requires temporarily and selectively unlocking the magnetic configuration of the storage layer alone. The use of different layer thicknesses AFM1(15 nm) and AFM2(6 nm) results in different Néel temperatures [1]. To experimentally verify the values of these temperatures, the following experiment was carried out. The film was examined using VSM equipped with a thermal setup. The sample was heated to a certain temperature and then cooled in the presence of +50 mT field. Then, a hysteresis loop was recorded at room temperature. Then the sample was then heated to the same temperature, but cooling occurred in the opposite field -50 mT. Then room temperature loop was again recorded. Fig.S2 shows a comparison of pairs of loops obtained at heating temperatures in the range from 120 to 240 C. Based on the loop from Fig.S1b, it can be noted that a complete switch of magnetization orientation in the storage layer occurs at a temperature of 180 C. Based on these indirect data, it can be concluded that the Néel temperature of the AFM2 layer is comparable to this value. A complete switch of magnetization orientation in the reference layer is observed at a temperature of 240 C.

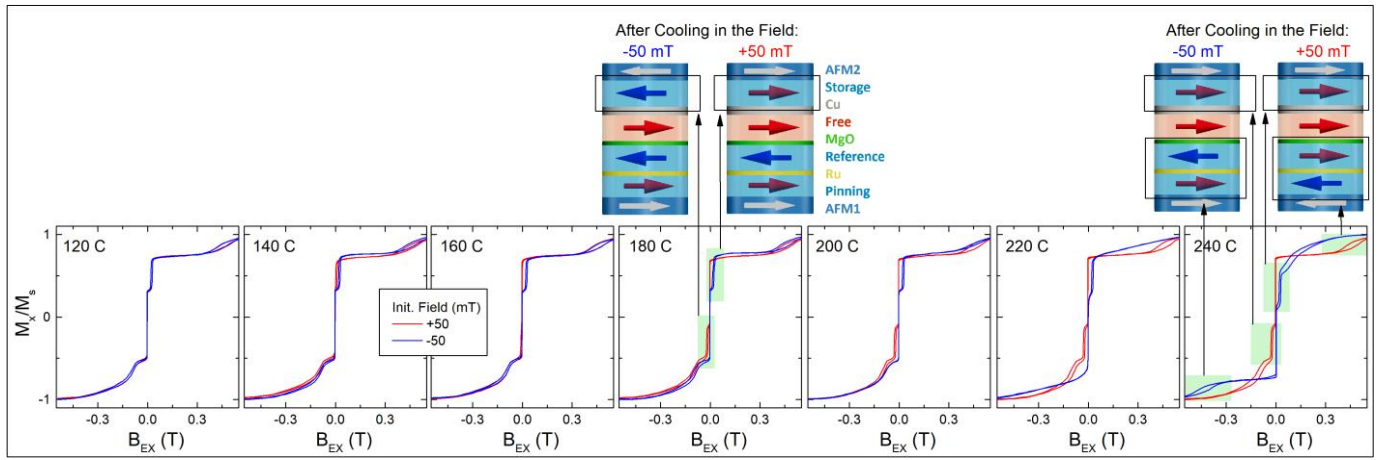

*Fig.S2. Hysteresis loops of* IrMn(15)/CoFe(2)/Ru(0.8)/CoFeB(2.6)/MgO/CoFeB(2)/Ta(0.2)/NiFe(7)/Ta(10)/NiFe(6)/IrMn(6) *films obtained by the VSM method at room temperature after annealing under various temperature for two different field orientation. Areas of orientation switching in the storage layer and reference layer are marked in green. The scheme shows orientation of remanent magnetization in the layers for the case of annealing at temperatures of 180 and 240 C when the orientation of the external field changes during cooling.*

S.3. Initialization pulse parameters

Magnetic reconfiguration of the structure is possible by applying a voltage pulse. In this case, as a result of Joule heating, the antiferromagnet layer in the storage layer is heated above the Neel temperature, which leads to the unlocking of the magnetic configuration in the adjacent ferromagnetic layer. The ferromagnetic layer is reconfigured in accordance with the magnitude and direction of the external magnetic field. Upon completion of the pulse, the ferromagnetic layer is frozen in its current state due to the cooling of the antiferromagnetic layer. The system contains two antiferromagnetic layers with different thicknesses and with different Neel temperatures: the bottom AFM1 layer is the pinning SAF layer and the upper AFM2 is the pinning storage layer. To control the state of the storage layer separately, it is necessary to define a voltage range that will unlock the storage layer selectively.

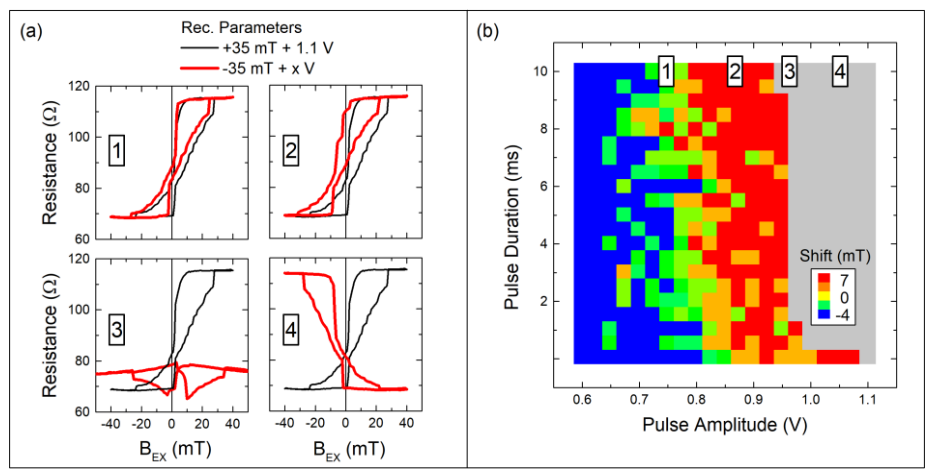

*Fig.S3. (a) Examples of changes in the hysteresis loop after a 1 ms pulse of variable amplitude, compared to the "base" state loop, obtained by action of the 1.1 V pulse in the presence of +35 mT. (b) The diagram describes the shift of the hysteresis loop depending on the duration and amplitude of the voltage pulse applied to the structure in "base" state. The grey area corresponds to the reconfiguration of the reference layer.*

To determine the required amplitude and duration of the pulse, the change in the process of magnetization reversal of the structure was assessed using a hysteresis loop. Before each pulse, the system was brought to the "basic" state: the reference and storage layers were magnetized in the +x direction, which was achieved by the pulse 1.1 V in the presence of a field +35 mT. Next, a pulse with an amplitude from 0.5 to 1.2 V and duration from 0.1 to 10 ms was applied in the presence of the opposite field -35 mT. Depending on the pulse parameters, four results can be distinguished: 1) the loop receives some intermediate displacement; 2) the loop receives maximum displacement; 3) the reference SAF loses the reference direction; 4) magnetization reversal of the reference SAF occurs, which causes the loop to become mirrored, as shown in Fig.S3a. The Fig.S3b shows a diagram illustrating the dependence of the loop shift on the pulse parameters. The area corresponding to the reversal of the SAF reference direction is marked in gray.

*S.4. Comparison of cases with different magnetization orientations in the reference layer*

Ideally, the dynamic properties of the device should depend only on the magnetic configuration of the storage layer for a given geometry and initialization parameters. However, the free layer also suffers an influence form the reference SAF. This interaction may include magnetostatic, indirect exchange or Néel coupling and is described in terms of the "effective magnetic field" equal to ±1.5 mT. The orientation of this field depends on the orientation of the magnetization in the pinning layer, which is magnetized opposite to the reference layer. A change in the orientation of the magnetization in the reference layer can occur as a result of reconfiguration, under the action of a high voltage pulse in the presence of a field (S.3). A qualitative difference of such switching is the change in the hysteresis loop transversal direction.

The Fig.S4a-b shows examples of the maximum shift of loops in the case of +x (State 1) and –x (State 2) magnetization orientation in the pinning layer. The Fig.S4c shows the dependence of the shift on the magnitude of the initializing field for these two cases. As can be seen, the dependences are shifted by ±1.5 mT for different magnetization orientations in the pinning layer. Since vortex oscillations cannot be excited in the central part of the disk, their presence is determined by the combination of the effective field and the magnetostic field from the storage layer. The Fig.S4d-e shows frequency diagrams depending on the initializing field for different magnetization orientations in the pinning layer. As can be seen, the region of excitation of oscillations changes to the opposite direction of $B_{init}$. Fig.S4f shows the dependences of the resonant frequencies for these two cases.

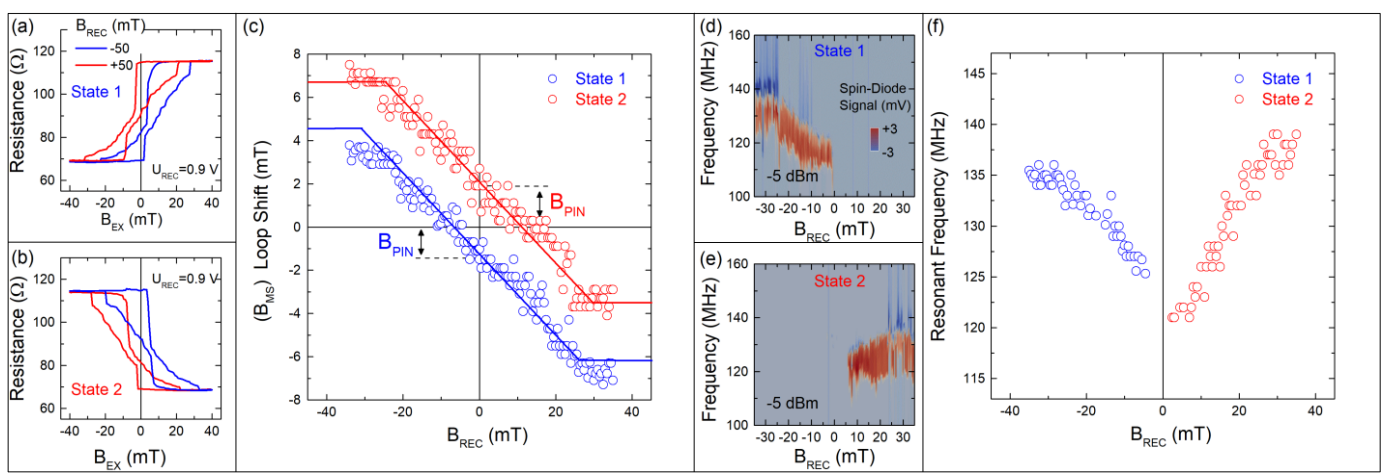

*Fig.S4. (a-b) Examples of hysteresis loops with maximum mutual shift for the cases of opposite magnetization orientation in the reference layer. (c) Dependence of the shift on the magnitude and direction of the initializing field due to reconfiguration of the storage layer. (d-e) Frequency diagrams for cases of different magnetization*

*orientations in the reference layer. (c) The general dependence of the resonant frequency on the magnitude and direction of the initializing field.*

S.5. Effect of Mutual Chirality of Vortexes

During the reconfiguration of the storage layer, a positive voltage pulse $U_{REC}$ is applied. The resulting electric current creates an Oersted field, the direction of which determines the chirality of the magnetic vortex in the storage layer. Let's take it to be clockwise - CW. In most cases, the same pulse also determines the CW chirality of the magnetic vortex in the free layer. However, the chirality of the free layer vortex can be set separately after reconfugarization. To do this, the storage layer is transferred to a single-domain state using an external field, and then the field is removed in the presence of a small direct current through the MTJ. By nucleation a vortex in the free layer in the presence of a current of +5 or -5 mA, chirality can be set to CW or CCW, illustration in the Fig.S5.

The diagrams in the Fig.S5a-b show the experimentally obtained dependences of the changes in the spectra for two combinations of vortex chiralities. The corresponding dependences of the resonant frequencies are combined in Fig.S5c. It can be seen that there is a slight quantitative difference, but the general trend of a smooth decrease in the resonant frequency remains.

Micromagnetic modeling was also carried out for the cases of a combination of vortices with the same and different chiralities in the storage layer and the free layer, Fig.S5f-g. The combined graph Fig.S5d shows that a change in chirality leads to the same changes that were observed experimentally. This conclusion concerns dependence in negative fields. In positive fields the experiment failed to evaluate the dependence. The simulation results suggest that this may be partly due to a decrease in the amplitude of the oscillations as seen in the Fig.S5e.

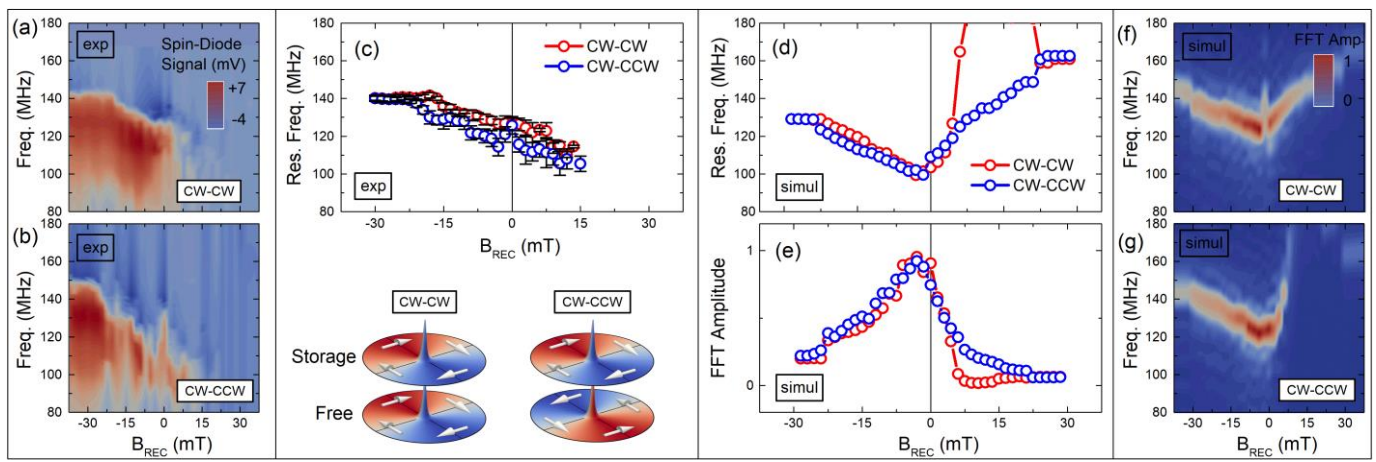

*Fig.S5. Experimentally obtained spectral diagrams depending on the magnitude of the reconfiguration field in the case of the same (a) and different (b) chirality of vortices in the storage and free layer. (c) Combined dependences of resonance frequencies for different combinations of chiralities. The corresponding dependencies (d) and diagrams (f-g) obtained by micromagnetic simulations methods are given for comparison. (e) Dependence of the oscillation amplitude on the reconfiguration field.*

S.6. Reproducibility considerations

The work investigated the behaviour of the loop shift and the resonance frequency of the vortex precession on the initialization parameters. To construct these dependences, current pulses were passed through the structure many hundreds of times in combination with an external field. The operating state of the structure is the presence of a magnetic vortex in the free layer. It should be noted that in some cases the vortex did not nucleate in the disk, which means that the shape of the hysteresis loop changed and the precession could not be excited. The Fig.S6a. shows examples of several qualitatively different loops. If, during the transition from a positive field to a negative one, and then from a negative to a positive one, a vortex is nucleated in the free layer, then this corresponds to loop #1. In case #2, when moving from a positive field to a negative field, the decrease in resistance occurs smoothly, and not abruptly as in case #1. This indicates a long-term existence of the c or s state. Then the vortex nucleate, but exists in a smaller range of fields. In case #3, during the transition from a positive field to a negative field, the vortex does not nucleate at all. In all cases other than #1, the loop shift estimate will fall out of the general dependence, because qualitative changes in the magnetization reversal process occur.

Even if a vortex exists in the free layer, the resonance curves may also have qualitative differences Fig.S6b. For example, the precession of the nucleus at the edge can lead to the exit of the nucleus beyond the disk, as in case #1. Due to the peculiarities of the method of excitation (spin-torque diode effect) of oscillations, it is problematic in the central region and the amplitude will be minimal #2. The existence of vortices with different chiralities is also possible, which leads to a reversal of the resonance curve #2-4 [2]. In all cases, the difference between the curve and case #2 will lead to a departure of the resonance frequency value from the general trend.

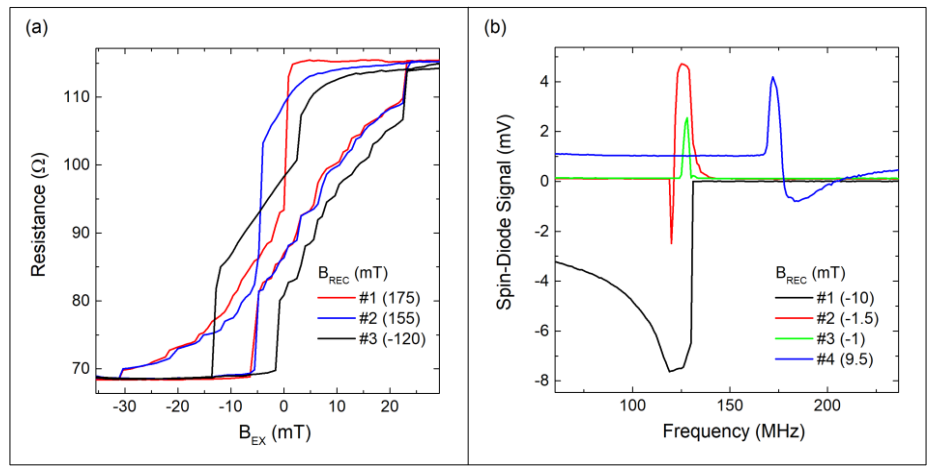

Fig.S6. (a) Examples of qualitatively different hysteresis loops. When the field changes from positive to negative: 1 - the vortex nucleates abruptly; 2 – the vortex nucleates through the c or s state; 3 – the vortex does not nucleate. (b) Examples of different spectra: 1 - the vortex annihilated during the process of oscillations at the edge of the disk; 2 and 4 - different chirality of the vortex; 3 - no change in signal sign.

*S.7. Simulation dependence of magnetostatic field on disk*

In the experimental structure, the storage layer acts on the free layer through a magnetostatic field, assuming that the 10 nm layer excludes exchange interaction. To study this effect at the micromagnetic level, the system was simulated using MuMax3 software. The program code is given in S.8. To improve the simulation efficiency, the behaviour of the storage disk from the external field was separately simulated, and then the magnetostatic field it would create in the free layer was calculated and stored. The magnetic structure of the free layer then relaxes in the presence of this field. Fig.7a. shows an example of the distribution of the magnetostatic field in the case of a single-domain state of the storage disk. Dependences of the magnetostatic field components along the central x-axis are shown in the Fig.S7b. It can be seen that even in the case of a single-domain configuration, the field is highly inhomogeneous. The x-component of the magnetostatic field decreases by more than half when moving from the periphery of the disk to the centre. In this case, the maximum z-component is three times the maximum x-component. The figure shows the dependence of the magnetostatic field component when changing the parameters of the storage disk: size and saturation magnetization, Fig.S7c. Reducing the disk diameter from 600 to 300 nm leads to an increase in the field component in the center by more than two times. Increasing the saturation magnetization of the disk material makes it possible to linearly increase the magnetostatic field.

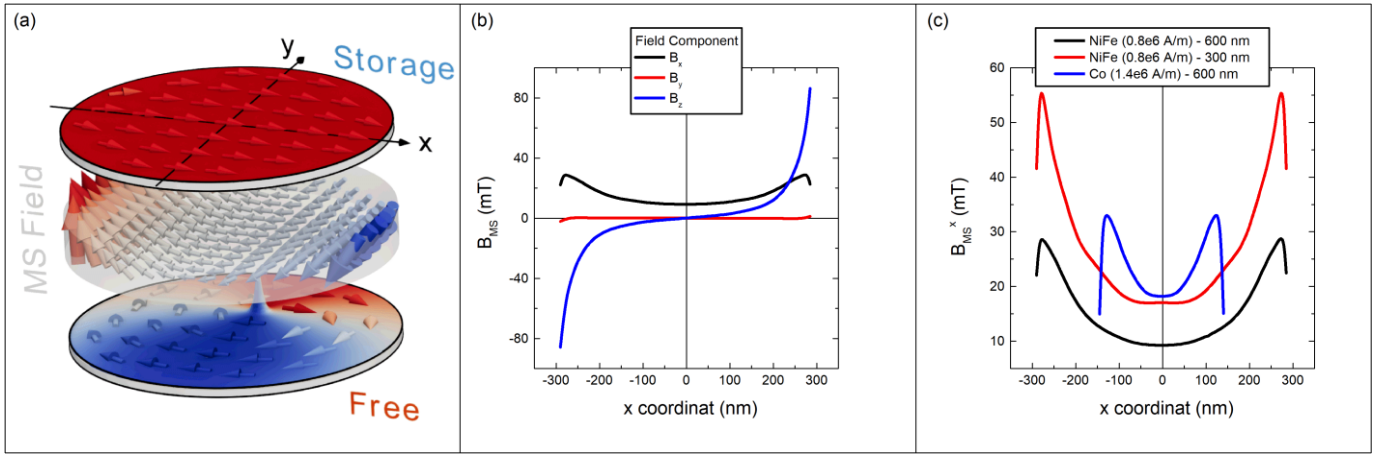

*Fig.S7. Results of micromagnetic simulation. (a) An example of the magnetic structure of a free layer for the case of freezing single domain in the storage layer. The grey cylinder contains the distribution of the magnetostatic field created by the storage layer in the free layer. The (b) plot shows the dependences of the components of the magnetostatic field in the free layer along the central axis for this case. (c) Dependence of the x-component distribution of the magnetostatic field in the free layer on magnetic parameters and size of storage disk.*

*S.8. Spectrum simulation*

To determine the influence of the magnetostatic field on the frequency of the gyrotropic motion of the vortex, the following method was used. Using MuMax3 [3], a disk with a diameter of 600 nm and a thickness of 6 nm with a cell of 5x5x6 nm was specified. The program code is given in S.10. The magnetic structure of the disk relaxed in the presence of a previously obtained magnetostatic field, which would create a storage layer in the free layer, and additional small field $B_y = 3\ mT$, Fig.S8a. Then this additional field was turned off and the relaxation of the system in time was considered with recording of magnetization components, Fig.S8b. The time step was 1 ps, and the total simulation time was 500 ns, which corresponds to a frequency resolution of 2 MHz. Decaying oscillations of magnetization, corresponding to oscillations of the vortex core, were processed using the fast Fourier transform method (FFT), which made it possible to obtain a spectrum, Fig.S8c. This method is well known and widely used [4] [37].

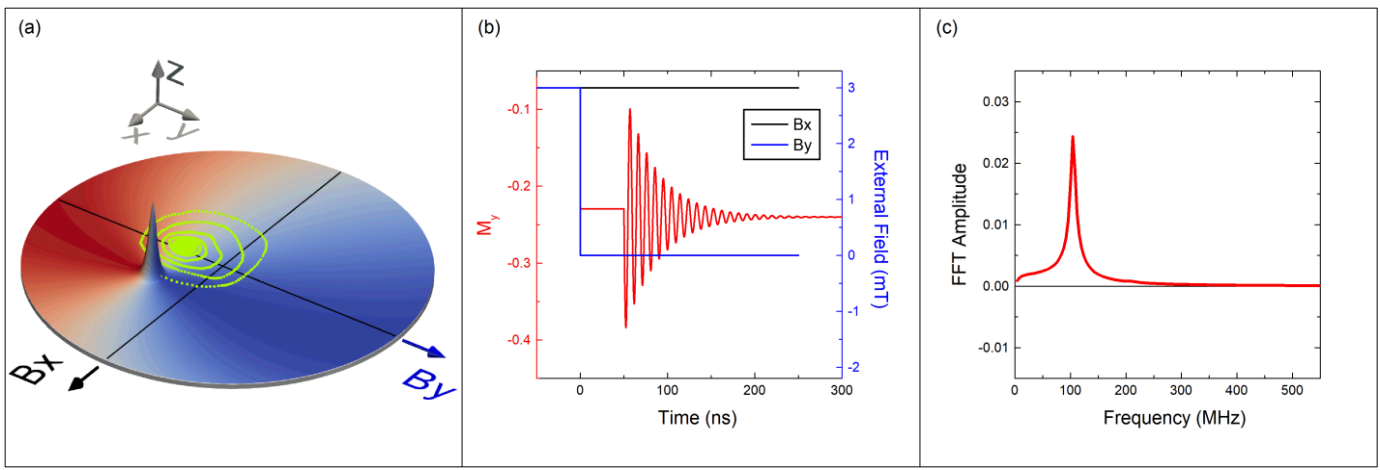

*Fig.S8. (a) The trajectory of the vortex core during the process of relaxation from a displaced state. (b) Dependence of the y-component of the magnetization on time during the relaxation process after turning off the y-field. (c) The spectrum obtained as a result of FFT transformation of the oscillating dependence of magnetization.*

*S.9. The influence of an external field on spectral characteristics*

Micromagnetic simulation was also carried out for the case of exposure to an external field on the free layer. To minimize the impact of the storage layer, the magnetic vortex in it was frozen in the center. A diagram was obtained depending on the magnitude of the constant external field, Fig.S9b. As can be seen, in the region of positive fields, the simulation results are in good agreement with the experimental results, Fig.S9c-d.

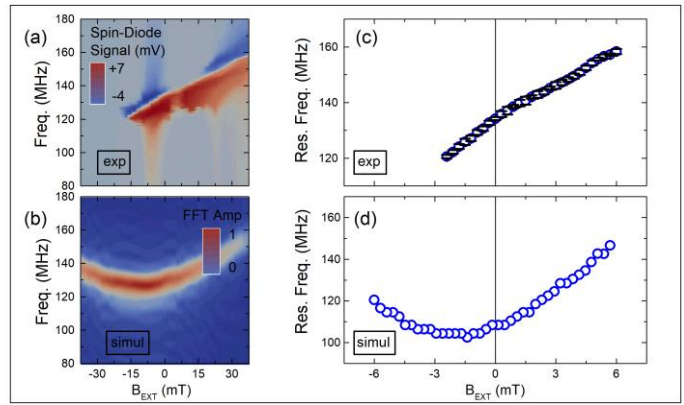

*Fig.S9. Experimentally measured spectral dependence (a) on a constant external field when a symmetrical vortex is frozen in the storage layer and the corresponding dependence of the resonance frequency (c). For the same conditions, the diagram (b) and dependence of the resonant frequency (d) obtained by micromagnetic simulation.*

*S.10. Simulation codes for MuMax3*

This section provides program codes for estimating the resonant frequency of the gyrotropic motion of a magnetic vortex in the free layer disk depending on the magnetic configuration of the storage layer disk. The task is divided into two parts. First, the disk of the storage layer was considered separately, Tab.1. The magnetic structure of the disk relaxed in the presence of an external field, performing the function of field $B_{REC}$ in the experiment. The external field varied in a range of ±32 mT, divided into one hundred steps. After each relaxation, a separate file was saved with the distribution of the magnetostatic field created by the disk under consideration at a distance corresponding to the position of the free disk.

Further, the disk of the free layer was considered separately, Tab.2. Its magnetic structure is minimized in the presence of the pre-stored magnetic field of the storage disk. After this, the process of damped oscillations is considered, as described in S.7 to determine the resonant frequency. This algorithm is repeated for each of the hundred pre-saved field distribution files.

*Table 1.*

| | 1_Magnetostatic_field.mx3 | |
|---|---|---|
| 1 | i:=0 | |
| 2 | SetGridsize(128, 128, 3) | |
| 3 | SetCellsize(5e-9, 5e-9, 6e-9) | |
| 4 | setgeom( circle(600e-9)) | |
| 5 | tableaddvar(i,"i","") | |
| 6 | TableAdd(B_ext) | |
| 7 | TableAdd(CropLayer(B_demag, 2)) | |
| 8 | OutputFormat = OVF2_TEXT | |
| 9 | defregion(0, layer(0)) | The storage disk is located in layer zero. To determine the magnetostatic field under the storage layer disk, empty space is specified. It is described by additional two layers. |
| 10 | defregion(1, layer(1)) | |
| 11 | defregion(2, layer(2)) | |
| 12 | Msat.setregion(0, 0.8e6) | The program does not allow voids, so zero parameters must be set in all layers. |
| 13 | Msat.setregion(1, 0e6) | |
| 14 | Msat.setregion(2, 0e6) | |
| 15 | Aex.setregion(0, 13e-12) | |

| 16 | Aex.setregion(1, 0e-12) | |
| 17 | Aex.setregion(2, 0e-12) | |
| 18 | frozenspins.SetRegion(1, 1) | Despite the zero magnetic moment, the program will perform calculations in all layers. To exclude this, the magnetic structure of the layers describing the void must be frozen. |
| 19 | frozenspins.SetRegion(2, 1) | |
| 20 | for i=0; i<101; i+=1{ | |
| 21 | print(i) | |
| 22 | m.setRegion(0, vortex(1,1)) | |
| 23 | m.setRegion(1, vortex(1,1)) | |
| 24 | m.setRegion(2, vortex(1,1)) | |
| 25 | B_ext = vector(-0.032*(1-i*2/100), 0, 0) | |
| 26 | print(B_ext) | |
| 27 | relax() | |
| 28 | SaveAs(CropLayer(B_demag, 2), sprintf("%d.ovf", i)) | The distribution of the magnetostatic field in the outermost layer is saved to a file. |
| 29 | save(m) | |
| 30 | tablesave() | |
| 31 | } | |

Table 2.

| | 2_ Resonance_frequency.mx3 | |
|---|---|---|
| 1 | i:=0 | |
| 2 | TableAdd(ext_corepos) | |
| 3 | tableaddvar(i,"i","") | |
| 4 | TableAdd(B_ext) | |
| 5 | OutputFormat = OVF2_TEXT | |
| 6 | SetGridsize(128, 128, 1) | |
| 7 | SetCellsize(5e-9, 5e-9, 6e-9) | |
| 8 | setgeom(circle(600e-9)) | |
| 9 | Msat = 800e3 | |
| 10 | Aex = 13e-12 | |
| 11 | alpha = 0.02 | |
| 12 | for i=15; i<101; i++ { | |
| 13 | B_ext.RemoveExtraTerms() | |
| 14 | m=vortex(-1,1) | |
| 15 | file_name:=sprintf("Magnetostatic_field.out//%d.ovf",i) | One by one, files with the distribution of the magnetostatic field are loaded from the folder with the results of the previous program. Magnetization is minimized in the presence of three fields: the magnetic field from the file; constant effective field from the pinning layer -2; bias field 3 mT. After minimization, the bias field is instantly turned off and the relaxation process of the system is recorded over time. |
| 16 | print(file_name) | |
| 17 | B_Ext.add(loadfile(file_name), 1) | |
| 18 | B_ext = vector(-0.002, 0.003, 0) | |
| 19 | relax() | |
| 20 | save(m) | |
| 21 | B_ext = vector(-0.002, 0, 0) | |
| 22 | run(500e-9) | |
| 23 | tableautosave(0.1e-9) | |
| 24 | } | |

# Bibliography

[1] L. Lombard, E. Gapihan, R. C. Sousa, Y. Dahmane, Y. Conraux, C. Portemont, C. Ducruet, C. Papusoi, I. L. Prejbeanu, J. P. Nozières, B. Dieny, A. Schuhl, "IrMn and FeMn blocking temperature dependence on heating pulse width," *Journal Of Applied Physics,* vol. 107, p. 09D728, 2010.

[2] L. Martins, A. Jenkins, L. Alvarez, P. Freitas, R. Ferreira, "Non-volatile artificial synapse based on a vortex nano-oscillator," *Scientific Reports,* vol. 11, no. 16094, 2021.

[3] Arne Vansteenkiste, Jonathan Leliaert, Mykola Dvornik, Mathias Helsen, Felipe Garcia-Sanchez, Bartel Van Waeyenberge, "The design and verification of MuMax3," *AIP Advances,* vol. 4, p. 107133, 2014.

[4] M. E. Stebliy, S. Jain, A. G. Kolesnikov, A. V. Ognev, A. S. Samardak, A. V. Davydenko, E. V. Sukovatitcina, L. A. Chebotkevich, J. Ding, J. Pearson, V. Khovaylo & V. Novosad, "Vortex dynamics and frequency splitting in vertically coupled nanomagnets," *Scientific Reports,* vol. 7, no. 1127, 2017.

[5] Gerardin, O., Gall, H., Vukadinovic, N. , "Micromagnetic calculation of the high frequency dynamics of nano-size rectangular ferromagnetic stripes," *J. Appl. Phys.,* vol. 89, p. 7012, 2001.